\font\bbb=msbm10                                                   

\def\C{\hbox{\bbb C}}

\def\R{\hbox{\bbb R}}

\def\JPA{{\sl J. Phys.\ A}}

\def\JSP{{\sl J. Stat.\ Phys.}}

\def\PA{{\sl Physica A}}

\def\PRD{{\sl Phys.\ Rev.\ D}}

\def\RMP{{\sl Rev.\ Mod.\ Phys.}}

\def\SPAWPM{{\sl Sitzungsber.\ Preuss.\ Akad.\ Wiss., Phys.-Math.}}

\def\ZPB{{\sl Z. Phys.\ B}}

\def\dajm{\hbox{D. A. Meyer}}

\def\albrecht{\hbox{A. Albrecht}}

\def\feynman{\hbox{R. P. Feynman}}

\def\zurek{\hbox{W. H. Zurek}}

\def\hfb{\hfil\break}

\catcode`@=11
\newskip\ttglue

   \font\ninerm=cmr9    \font\eightrm=cmr8   \font\sixrm=cmr6
  \font\ninebf=cmbx9   \font\eightbf=cmbx8  \font\sixbf=cmbx6
  \font\nineit=cmti9   \font\eightit=cmti8  
  \font\ninesl=cmsl9   \font\eightsl=cmsl8  
  \font\ninemi=cmmi9   \font\eightmi=cmmi8  \font\sixmi=cmmi6

\font\bigtenbf=cmr10 scaled\magstep2 

\def\ninepoint{\def\rm{\fam0\ninerm}%
  \textfont0=\ninerm \scriptfont0=\sixrm
  \textfont1=\ninemi \scriptfont1=\sixmi
  \textfont\itfam=\nineit  \def\it{\fam\itfam\nineit}%
  \textfont\slfam=\ninesl  \def\sl{\fam\slfam\ninesl}%
  \textfont\bffam=\ninebf  \scriptfont\bffam=\sixbf
    \def\bf{\fam\bffam\ninebf}%
  \tt \ttglue=.5em plus.25em minus.15em
  \normalbaselineskip=11pt
  \setbox\strutbox=\hbox{\vrule height8pt depth3pt width0pt}%
  \normalbaselines\rm}

\def\eightpoint{\def\rm{\fam0\eightrm}%
  \textfont0=\eightrm \scriptfont0=\sixrm
  \textfont1=\eightmi \scriptfont1=\sixmi
  \textfont\itfam=\eightit  \def\it{\fam\itfam\eightit}%
  \textfont\slfam=\eightsl  \def\sl{\fam\slfam\eightsl}%
  \textfont\bffam=\eightbf  \scriptfont\bffam=\sixbf
    \def\bf{\fam\bffam\eightbf}%
  \tt \ttglue=.5em plus.25em minus.15em
  \normalbaselineskip=9pt
  \setbox\strutbox=\hbox{\vrule height7pt depth2pt width0pt}%
  \normalbaselines\rm}

\def\sfootnote#1{\edef\@sf{\spacefactor\the\spacefactor}#1\@sf
      \insert\footins\bgroup\eightpoint
      \interlinepenalty100 \let\par=\endgraf
        \leftskip=0pt \rightskip=0pt
        \splittopskip=10pt plus 1pt minus 1pt \floatingpenalty=20000
        \parskip=0pt\smallskip\item{#1}\bgroup\strut\aftergroup\@foot\let\next}
\skip\footins=12pt plus 2pt minus 2pt
\dimen\footins=30pc

\def\ie{{\it i.e.}}

\def\etc{{\it etc.}}

\magnification=1200
\input epsf.tex

\dimen0=\hsize \divide\dimen0 by 13 \dimendef\chasm=0
\dimen1=\hsize \advance\dimen1 by -\chasm \dimendef\usewidth=1
\dimen2=\usewidth \divide\dimen2 by 2 \dimendef\halfwidth=2
\dimen3=\usewidth \divide\dimen3 by 3 \dimendef\thirdwidth=3
\dimen4=\hsize \advance\dimen4 by -\halfwidth \dimendef\secondstart=4
\dimen5=\halfwidth \advance\dimen5 by -10pt \dimendef\indenthalfwidth=5
\dimen6=\thirdwidth \multiply\dimen6 by 2 \dimendef\twothirdswidth=6
\dimen7=\twothirdswidth \divide\dimen7 by 4 \dimendef\qttw=7
\dimen8=\qttw \divide\dimen8 by 4 \dimendef\qqttw=8
\dimen9=\qqttw \divide\dimen9 by 4 \dimendef\qqqttw=9

\parskip=0pt\parindent=0pt

\line{\hfil December 1996}
\line{\hfil quant-ph/9804023}
\bigskip\bigskip\bigskip
\centerline{\bf\bigtenbf DECOHERENCE IN THE DIRAC EQUATION}
\vfill
\centerline{\bf David A. Meyer}
\bigskip 
\centerline{\sl Project in Geometry and Physics}
\centerline{\sl Department of Mathematics}
\centerline{\sl University of California/San Diego}
\centerline{\sl La Jolla, CA 92093-0112}
\centerline{dmeyer@chonji.ucsd.edu}
\smallskip
\centerline{\sl and}
\smallskip
\centerline{\sl Institute for Physical Sciences}
\centerline{\sl Los Alamos, NM}
\vfill
\centerline{ABSTRACT}
\bigskip
\noindent A Dirac particle is represented by a unitarily evolving 
state vector in a Hilbert space which factors as 
$H_{\rm spin} \otimes H_{\rm position}$.  Motivated by the similarity
to simple models of decoherence consisting of a two state system 
coupled to an environment, we investigate the occurence of decoherence
in the Dirac equation upon tracing over position.  We conclude that 
the physics of this mathematically exact model for decoherence is 
closely related to {\it Zitterbewegung}.

\bigskip
\global\setbox1=\hbox{PACS numbers:\enspace}
\global\setbox2=\hbox{PACS numbers:}
\parindent=\wd1
\item{PACS numbers:}  03.65.Bz,  
                      03.65.Pm,  
                      05.30.-d.  
\item{\hbox to \wd2{KEY\hfill WORDS:}}   
                      decoherence; Dirac equation; entropy; 
                      {\it Zitterbewegung}.

\vfill
\eject

\headline{\ninepoint\it Decoherence in the Dirac equation
          \hfil David A. Meyer}
\parskip=10pt
\parindent=20pt

Decoherence---the process by which a quantum system initially in a
pure state evolves into a mixed state as it interacts with its 
environment---has been studied both in toy models designed to 
elucidate aspects of quantum measurement [1--4] and in more complex 
ones modelling realistic physical situations [3,5--7].  In the purely 
quantum mechanical models the Hilbert space factors as a tensor 
product $H = H_{\rm sys} \otimes H_{\rm env}$ of Hilbert spaces 
describing the degrees of freedom of the system and environment,
respectively.  From the complete state $|\Psi\rangle \in H$ one forms
the density operator $\rho := |\Psi\rangle \otimes \langle\Psi|$, 
which is traced over $H_{\rm env}$ to give the reduced density 
operator $\tilde\rho := {\rm Tr}_{\rm env}\rho$.  Physically, the 
trace corresponds to the absence of measurements on the environment; 
$\tilde\rho$ is the average over all possible states thereof.  
Generically, $\tilde\rho$ is not of the form 
$|\Psi_{\rm sys}\rangle \otimes \langle\Psi_{\rm sys}|$ for any
$|\Psi_{\rm sys}\rangle \in H_{\rm sys}$; it is rather a linear
combination $\sum p_i |\psi_i\rangle \otimes \langle\psi_i|$, 
where $|\psi_i\rangle \in H_{\rm sys}$ and the coefficients $p_i$ are
positive and sum to 1.  That is, $\tilde\rho$ describes a mixed rather 
than a pure state, the degree of mixing being measured by the entropy
$S := -{\rm Tr}\tilde\rho \log\tilde\rho$.

The simplest decoherence models considered typically have a two 
dimensional $H_{\rm sys}$ coupled to an environment whose Hilbert 
space is also two [1--3] or at most finite [2,4] dimensional, although 
more realistic (infinite dimensional) environments have been used to 
investigate the absence of chiral superpositions in molecules [3], for
example.  Now, the Dirac equation describes the unitary evolution of a 
spin-${1 \over 2}$ particle---the Hilbert space factors as 
$H = H_{\rm spin} \otimes H_{\rm position}$---and $H_{\rm spin}$ is
finite dimensional.  In the spinor chain path integral formulation of 
the Dirac equation [8,9], to each direction in space is associated a 
spin projection, and over a time interval $\delta$ a projection 
operator is understood to propagate the particle a distance $\delta$ 
in the associated direction.  A single path contributing to a 
transition amplitude from $x$ to $x'$ consists of a sequence of spin 
projections and chiralities such that the sum of the translations 
associated with the spin projections, each multiplied by the 
corresponding chirality, is proportional to the difference in position 
$x' - x$, where the proportionality constant depends on the spatial 
dimensionality.  The amplitude of such a path is proportional to 
$(i\delta m)^r$ (in units with $\hbar \equiv 1 \equiv c$), where $m$ 
is the mass of the particle and $r$ is the number of chirality 
reversals.  

{}From the point of view of decoherence models, this path integral
describes exactly how the spin degrees of freedom are `recorded' by
the position of the particle.  That is, $H_{\rm position}$ can be 
thought of as the Hilbert space of an environment measuring a quantum
system with Hilbert space $H_{\rm spin}$.  Thus the Dirac equation can 
be analyzed for decoherence {\it via\/} reduced density matrices.  In
$1+1$ dimensions%
\sfootnote*{Since there is no spin---only chirality---in $1+1$ 
            dimensions, this will simplify the subsequent discussion
            with no conceptual loss and allow direct comparison with
            decoherence of other two state systems [1--4].}
the particle wave function is a two component spinor field 
$\psi(x,t) = \bigl(\psi_{-1}(x,t),\psi_{+1}(x,t)\bigr)$ in the chiral
basis, so the density function is
$\rho(x,\alpha,x',\alpha',t) 
 = \psi_{\alpha}(x,t) \overline{\psi_{\alpha'}(x',t)}$, for 
$\alpha,\alpha' \in \{\pm1\}$.  Tracing over the environment is here 
an integration over position, so the reduced density matrix has four
components
$$
\tilde\rho_{\alpha\alpha'}(t)
 = \int\psi_{\alpha}(x,t)\overline{\psi_{\alpha'}(x,t)}\,dx.  \eqno(1)
$$              
In this note we examine (1) for decoherence by computing, exactly, its
entropy as a function of time.

Using the path integral formulation of the Dirac equation it is 
straightforward to compute the propagator 
$K_{\alpha_2\alpha_1}(x_2,t_2,x_1,t_1)$ satisfying
$$
\psi_{\alpha_2}(x_2,t_2) 
 = \int_{x_2-\Delta t}^{x_2+\Delta t}\! 
    K_{\alpha_2\alpha_1}(x_2,t_2,x_1,t_1) 
    \psi_{\alpha_1}(x_1,t_1)\,dx_1,                           \eqno(2)
$$
where $\Delta t := t_2 - t_1$.  The result is [8,9,10]:
$$
K_{\alpha_2\alpha_1}(x_2,t_2,x_1,t_1)
 = \cases{ \delta(\Delta t -\alpha_1\Delta x) 
                -(\Delta t +\alpha_1\Delta x) m J_1(m\tau) / 2\tau
                                   & if $\alpha_2     = \alpha_1$; \cr
           i m J_0(m\tau) / 2      & if $\alpha_2 \not= \alpha_1$, \cr
         }                                                    \eqno(3)
$$
where $\Delta x := x_2 - x_1$, 
$\tau := \sqrt{(\Delta t)^2 - (\Delta x)^2}$ and $J_i$ denotes the 
$i^{\rm th}$ order Bessel function of the first kind.  Using (2) and 
(3) in (1) gives an exact formula for the reduced density matrix as a
function of time in terms of the initial state of the particle.

\moveright\secondstart\vtop to 0pt{\hsize=\halfwidth
\null
\vskip -1.25\baselineskip   
$$
\epsfxsize=\halfwidth\epsfbox{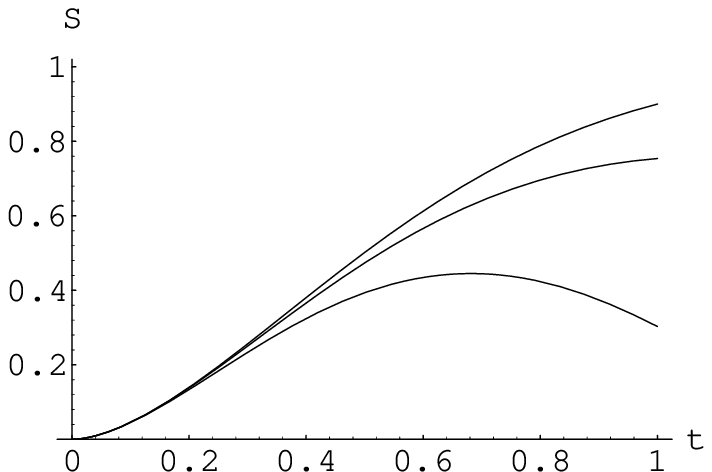}
$$
\vskip -0.5\baselineskip
\eightpoint{%
\noindent{\bf Figure~1}.  Entropy as a function of time for the 
reduced density matrix (1) with the particle in an initial equal 
superposition of chiralities and localized in a Gaussian around
$x = 0$.  From top to bottom, the curves are for increasing mass.
}}
\vskip-\baselineskip
\parshape=17
0pt \halfwidth
0pt \halfwidth
0pt \halfwidth
0pt \halfwidth
0pt \halfwidth
0pt \halfwidth
0pt \halfwidth
0pt \halfwidth
0pt \halfwidth
0pt \halfwidth
0pt \halfwidth
0pt \halfwidth
0pt \halfwidth
0pt \halfwidth
0pt \halfwidth
0pt \halfwidth
0pt \hsize
Decoherence is usually studied for an initial state which is a tensor 
product pure state 
$|\Psi_{\rm sys}\rangle \otimes |\Psi_{\rm env}\rangle$ [1--7], so 
consider the case $\psi(x,0) = f(x) {1 \choose 1}$, where 
$f(x)$ is proportional to $e^{-x^2/2}$ and is normalized so that 
$\int\psi^{\dagger}(x,0) \psi(x,0)\,dx = 1$.  That is, the particle is 
initially in an equal superposition of negative and positive 
chiralities and is localized near $x = 0$.  Figure~1 shows the entropy 
$S(t)$ of the reduced density matrix (1) for the range of masses 
$m \in \{0,1,2\}$.  Near $t = 0$ the rate of entropy gain increases as
the mass decreases:  the top curve is for $m = 0$ while the bottom one 
is for $m = 2$.  Thus, as could have been anticipated from the 
amplitudes associated to paths in the spinor chain path integral, 
$m^{-1}$ controls the strength of the system/environment interaction:  
when $m^{-1}$ is larger, decoherence occurs more rapidly.

To understand this decoherence, we begin by considering $m = 0$.  In 
this case the Bessel function terms in (3) vanish so the effect of the 
propagator is simply to translate the initial chirality position 
amplitudes along the lightcone:  
$\psi(x,t) = f(x + t) {1 \choose 0} + f(x - t) {0 \choose 1}$.  For 
the initial position Gaussian we can compute the reduced density 
matrix (1) exactly to get:
$$
\tilde\rho(t) = {1 \over 2}\pmatrix{       1 & e^{-t^2} \cr
                                    e^{-t^2} &        1 \cr}. \eqno(4)
$$
The diagonal terms in (4) are constant and equal, while the 
off-diagonal terms go rapidly to 0; thus the entropy goes rapidly to 
1 as indicated by Figure~1.

Next, we consider $m = 1$.  Just as in the massless case, if we were
to plot the position distributions for each chirality at $t = 0$, \ie,
$|\psi_{-1}(x,0)|^2$ and $|\psi_{+1}(x,0)|^2$, they would be two
coincident Gaussians centered at $x = 0$; there is no initial 
correlation between position and chirality for a tensor product state.
Figures~2 and 3 show the chirality position distributions at $t = 1$
for $m = 0$ and $m = 1$, respectively.  In each case the distributions
separate:  a particle with negative/positive chirality tends to move
to the left/right.  Now chirality is correlated with position, more
strongly for $m = 0$ than for $m = 1$, and it is this correlation 
which is measured by the entropy upon tracing over position.  For 
$m = 1$ dispersion distorts the initial Gaussian so that there is a
weaker correlation between chirality and position than for $m = 0$, 
and thus lower entropy.

\topinsert
$$
\epsfxsize=\halfwidth\epsfbox{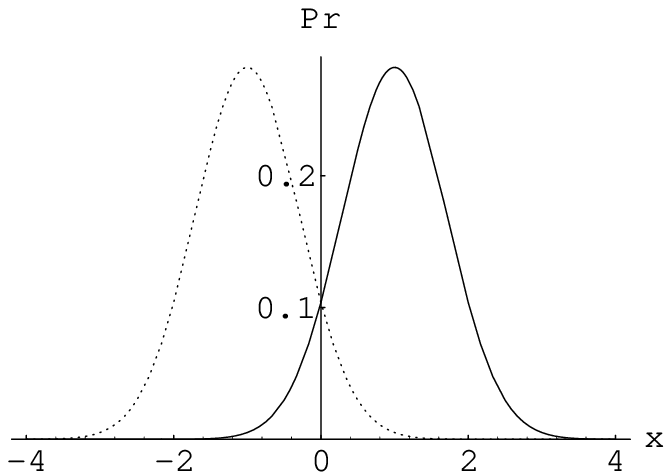}\hskip\chasm%
\epsfxsize=\halfwidth\epsfbox{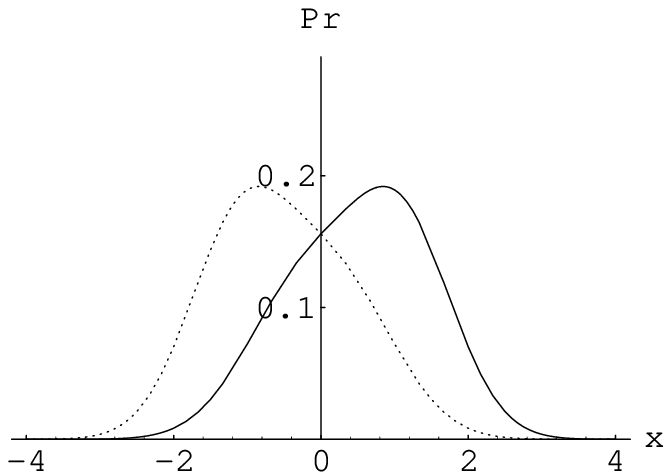}
$$
\hbox to\hsize{%
\vbox{\hsize=\halfwidth\eightpoint{%
\noindent{\bf Figure~2}.  The $t = 1$ chirality position distributions
for $m = 0$.  The distribution with the peak at the left/right is for 
negative/positive chirality; each remains a Gaussian.
}}
\hfill%
\vbox{\hsize=\halfwidth\eightpoint{%
\noindent{\bf Figure~3}.  The same chirality position distributions
shown in Figure~2 but for the case $m = 1$.  Dispersion distorts the
initial Gaussian and reduces the separation of the distributions.
}}}
\endinsert

Since the spin system is coupled to an infinite dimensional position
Hilbert space environment, not a small finite dimensional Hilbert 
space with a short Poincar\'e recurrence time [2--4], one might have 
expected even this weaker chirality/position correlation to continue 
to build and the entropy to increase asymptotically to 1 just as it
does for $m = 0$.  This expectation is contradicted, however, by the
$m = 2$ curve in Figure~1.  In this case the entropy reaches a (local) 
maximum and then begins to decrease.  In fact, the oscillatory 
behavior beginning here characterizes all of the massive entropy 
curves.

To explain how the entropy can decrease, despite the infinite 
dimensional position Hilbert space, Figure~4 shows the $m = 1$ entropy 
curve over a longer time interval $0 \le t \le 2$.  The inset graphs 
show the same chirality position distributions plotted in Figures~2
and 3, for times $t \in \{0.5,1,1.5,2\}$.  
The slowing and then reversal of the entropy increase is due to the 
decreasing portion of the distributions in the fastest moving
(outermost) peaks and the resulting increasing portion of each
distribution which overlaps the other.%
\sfootnote*{Spacetime plots of quantum lattice gas automaton 
simulations of the Dirac equation showing exactly this behavior 
(although with periodic spatial coordinate) can be found in [10].}
That is, dispersion acts initially to increase (by separating the 
distributions) and subsequently to decrease (by broadening/distorting
the distributions) entropy; then the cycle repeats.  For smaller 
masses not only does the initial increase occur faster, but the 
entropy also reaches greater values before starting to decrease.

\topinsert
$$
\epsfxsize=\usewidth\epsfbox{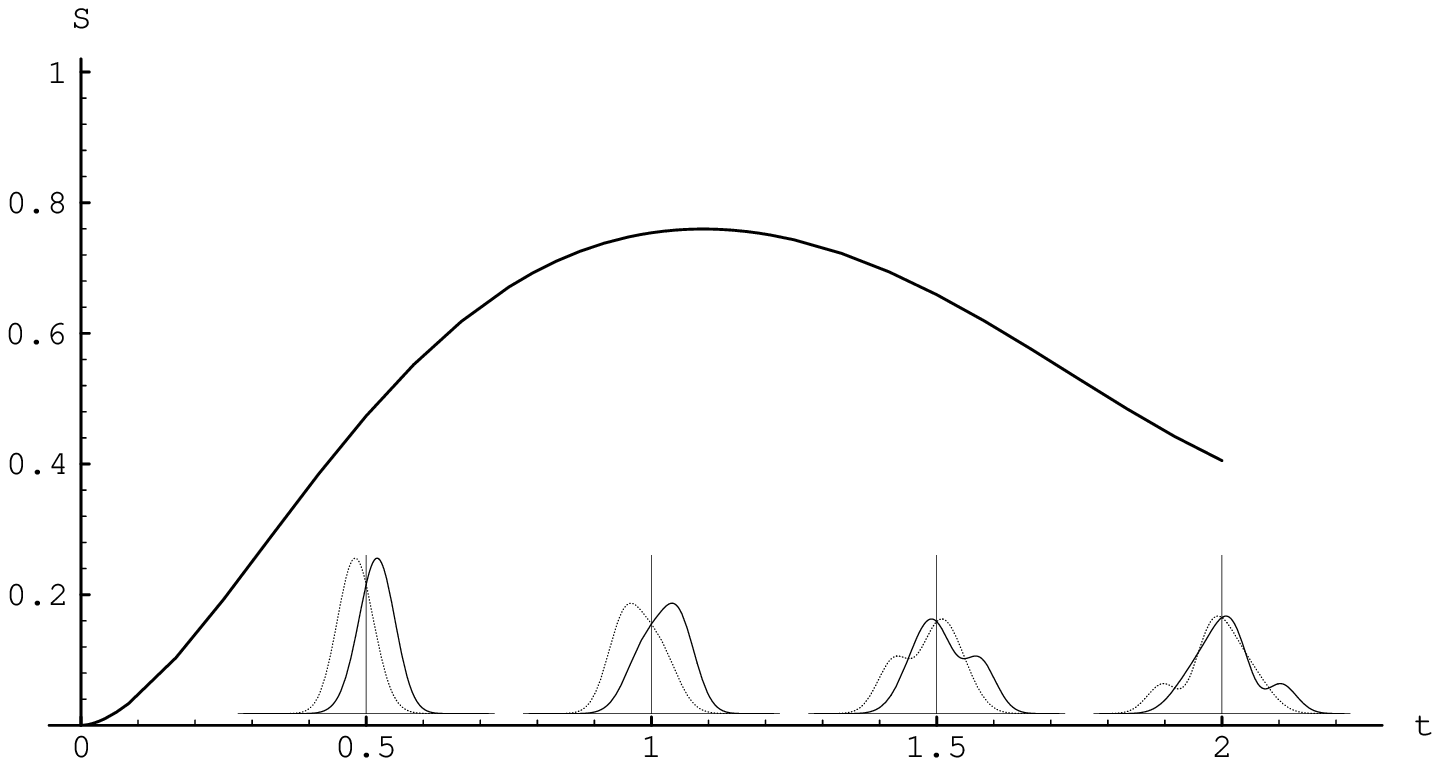}
$$
\eightpoint{%
{\narrower\noindent{\bf Figure~4}.  Entropy plotted as a function of 
time for an initial position Gaussian with $m = 1$.  The inset graphs 
show the chirality position distributions at half integer times.\par}
}
\endinsert

The results shown in Figures~1--4 have been obtained for quite 
special initial conditions:  in each case an equal superposition of
negative and positive chiralities.  While this choice demonstrates
decoherence in the Dirac equation particularly clearly, any spinor 
tensored with an initial position Gaussian will decohere.  Figure~5
shows the entropy curve over $0 \le t \le 1$ for initial condition
$\psi(x,0) \propto e^{-x^2/2} {0 \choose 1}$ with $m = 1$, while 
Figure~6 shows the chirality position distributions at $t = 0.5$.  The 
larger distribution is $|\psi_{+1}(x,0.5)|^2$; as expected, since the 
initial chirality is purely positive, the chirality is still more 
likely to be positive than negative at $t = 0.5$.  Nevertheless, there
is some probability that the particle\break
\vskip-\baselineskip

\midinsert
$$
\epsfxsize=\halfwidth\epsfbox{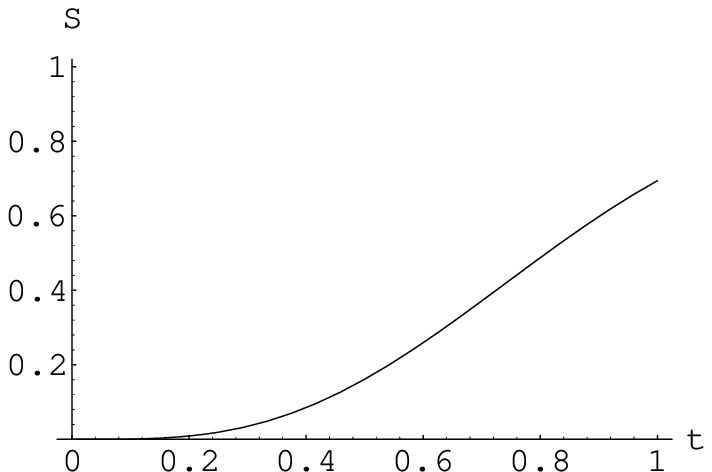}\hskip\chasm%
\epsfxsize=\halfwidth\epsfbox{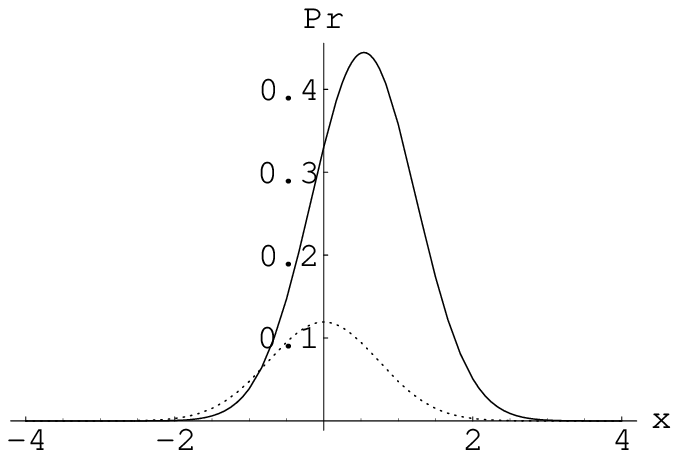}
$$
\hbox to\hsize{%
\vbox{\hsize=\halfwidth\eightpoint{%
\noindent{\bf Figure~5}.  Entropy as a function of time for the chiral
initial condition localized by a position Gaussian and for $m = 1$.
Initial entropy gain is slower than in the corresponding parity 
invariant case shown as the middle curve in Figure~1.
}}
\hfill%
\vbox{\hsize=\halfwidth\eightpoint{%
\noindent{\bf Figure~6}.  The distributions $|\psi_{\pm1}(x,0.5)|^2$
for the same initial condition as in Figure~5.  The larger
distribution is for positive chirality; the smaller, for negative.  
There is a small chirality/position correlation.
}}}
\endinsert

\noindent has negative chirality, so
chirality/position correlation has developed.  Just as in the parity 
invariant cases, only more slowly, this correlation causes the entropy 
increase shown in Figure~5.

By superposition in $H_{\rm sys}$, therefore, any tensor product 
initial condition of 
the form $\psi(x,0) \propto e^{-x^2/2} (a,b)$ decoheres.  This result, 
and its familiarity in the context of other models of decoherence 
[1--7], might lead us to believe that {\sl any\/} tensor product 
state, initially localized in position, decoheres---and to conclude 
that the spin of a Dirac particle evolves, absent position 
measurement, stochastically rather than purely quantum mechanically.
This, of course, is {\sl not\/} true, as we shall see by working in 
the stationary state basis.

The stationary states of the Dirac equation are the plane waves
$\psi^{(k,\epsilon)}(x) = e^{ikx} u(k,\epsilon)$, where $k \in \R$, 
$\epsilon \in \{\pm1\}$, and the fixed spinors 
$u(k,\epsilon) \in \C^2$ are normalized and orthogonal for 
$\epsilon = 1$ and $-1$.  While it is not localized in space, of 
course, suppose we take a plane wave $\psi^{(k,\epsilon)}$ as our
initial tensor product state.  Since this stationary state evolves by
multiplication by the phase $e^{-i\epsilon\omega t}$, where the energy
$\epsilon\omega$ satisfies the dispersion relation 
$\omega^2 = k^2 + m^2$, the density operator is constant:
$$
\rho(x,x',t) 
 =  e^{ik(x-x')} u(k,\epsilon) \otimes u^{\dagger}(k,\epsilon)
 =: e^{ik(x-x')} \tilde\rho(t),
$$
as is the reduced density matrix.  Thus stationary states do 
{\sl not\/} decohere, a familiar result in decoherence models with 
tensor product eigenstates of the Hamiltonian [1,3,4].

Any initial state can be expanded as a superposition of plane waves,
\ie, 
$$
\psi(x,0) 
 = \sum_{\epsilon} \int dk\, 
    \hat\psi(k,\epsilon) e^{ikx} u(k,\epsilon).
$$
In this form the time evolution is transparent:
$$
\psi(x,t) 
 = \sum_{\epsilon} \int dk\, 
    \hat\psi(k,\epsilon) e^{ikx-i\epsilon\omega t} u(k,\epsilon),
$$
so it is easy to write down the time dependent density operator 
explicitly:
$$
\rho(x,x',t)
 = \sum_{\epsilon,\epsilon'} \int dk dk'\,
    \hat\psi(k,\epsilon) e^{ikx-i\epsilon\omega t} 
    u(k,\epsilon) \otimes u^{\dagger}(k',\epsilon')
    e^{-ik'x'+i\epsilon'\omega't} \overline{\hat\psi(k',\epsilon')}.
                                                              \eqno(5)
$$
To obtain the time dependent reduced density matrix we need only set
$x' \equiv x$ in (5) and integrate over $x$.  The integral of
$e^{i(k-k')x}$ gives a delta function of $k - k'$, so we can evaluate
the integral over $k'$ to get
$$
\tilde\rho(t) 
 = \sum_{\epsilon,\epsilon'} \int dk\,
    \hat\psi(k,\epsilon) e^{-i\epsilon\omega t} 
    u(k,\epsilon) \otimes u^{\dagger}(k,\epsilon')
    e^{i\epsilon'\omega t} \overline{\hat\psi(k,\epsilon')}.  \eqno(6)
$$

The only nontrivial time dependence of $\tilde\rho(t)$ derives from 
terms in (6) with a factor of the form 
$e^{-i(\epsilon-\epsilon')\omega t}$ for $\epsilon' \not= \epsilon$.
Thus an initial state (whether a tensor product  or not) decoheres 
only if it has nonzero amplitudes $\hat\psi(k,\epsilon)$ and 
$\hat\psi(k,-\epsilon)$ for some $k \in \R$.  That is, decoherence
depends on the presence of positive and negative energy modes of the 
same momentum.  The initial states whose decoherence we exhibited in
Figures~1--5 are superpositions of such modes; in contrast, a wave 
packet constructed from only positive energy plane waves, would
disperse, of course, but not decohere.

We conclude by remarking that while the Dirac equation provides a 
mathematically exact model of decoherence, this decoherence is 
another aspect of the physical difficulties with the one particle 
interpretation of the Dirac equation.  The particle decoheres and 
behaves stochastically rather than quantum mechanically exactly to the 
extent that it is not identified as an electron or positron, for 
example.  The same interference between positive and negative energy 
states leads to {\it Zitterbewegung\/} [11].  That the entropy 
oscillates, rather than increasing permanently as it does in other
decoherence models with infinite dimensional environment Hilbert 
spaces [5--7], is a consequence of the same oscillating factors 
$e^{-2i\epsilon\omega t}$ in (6) that cause {\it Zitterbewegung}.  To
revise FitzGerald's rendering of {\sl The Rub\'aiy\'at} of Omar 
Khayy\'am [12]:

\medskip
\global\setbox3=\hbox{The Moving Finger writes; and, having writ,}
\centerline{The Moving Finger writes; and, having writ,}
\centerline{\hbox to \wd3{\it Zitterbewegt sich.\hfill}}

\vfill
\eject

\noindent{\bf Acknowledgements}
\nobreak

\nobreak
\noindent It is a pleasure to thank Mike Freedman, Brosl Hasslacher,
Melanie Quong, Jeff Rabin and Hans Wenzl for discussions about various 
aspects of this work.  I would also like to thank Sun Microsystems for 
providing support for the computational aspects of the project.

\bigskip

\global\setbox1=\hbox{[00]\enspace}
\parindent=\wd1

\noindent{\bf References}
\bigskip

\parskip=0pt
\item{[1]}
\zurek,
``Pointer basis of quantum apparatus:  Into what mixture does the wave
  packet collapse?'',
\PRD\ {\bf 24} (1981) 1516--1525.

\item{[2]}
\zurek,
``Environment-induced superselection rules'',
\PRD\ {\bf 26} (1982) 1862--1880.

\item{[3]}
E. Joos and H. D. Zeh,
``The emergence of classical properties through interaction with the
  environment'',
\ZPB\ {\bf 59} (1985) 223--243.

\item{[4]}
\albrecht,
``Investigating decoherence in a simple system'',
\PRD\ {\bf 46} (1992) 5504--5520;\hfb
\albrecht,
``Following a `collapsing' wave function'',
\PRD\ {\bf 48} (1993) 3768--3778.

\item{[5]}
A. O. Caldeira and A. J. Leggett,
``Path integral approach to quantum Brownian motion'',
\PA\ {\bf 121} (1983) 587--616.

\item{[6]}
W. G. Unruh and W. H. Zurek,
``Reduction of a wave packet in quantum Brownian motion'',
\PRD\ {\bf 40} (1989) 1071--1094.

\item{[7]}
J. P. Paz, S. Habib and W. H. Zurek,
``Reduction of the wave packet:  Preferred observables and 
  decoherence time scale'',
\PRD\ {\bf 47} (1993) 488--501.

\item{[8]}
\feynman\ and A. R. Hibbs,
{\sl Quantum Mechanics and Path Integrals}
(New York:  McGraw-Hill 1965);\hfb
S. S. Schweber,
``Feynman and the visualization of space-time processes'',
\RMP\ {\bf 58} (1986) 449--508.

\item{[9]}
T. Jacobson,
``Spinor chain path integral for the Dirac equation'',
\JPA\ {\bf 17} (1984) 2433--2451.

\item{[10]}
\dajm,
``From quantum cellular automata to quantum lattice gases'',
\JSP\ {\bf 85} (1996) 551--574.

\item{[11]}
E. Schr\"odinger,
``{\it \"Uber die kr\"aftefreie Bewegung in der relativistischen
       Quantenme\-chanik}'',
\SPAWPM\ {\bf XXIV} (1930) 418--428.

\item{[12]}
R. Arnot,
{\sl The Sufistic Quatrains of Omar Khayyam},
including tranlations of E. FitzGerald, \etc\
(NY:  M. Walter Dunne 1901);
in FitzGerald's Fifth Edition of {\sl The Rub\'aiy\'at},
Stanza {\it lxxi}.

\bye